\documentstyle[Sprocl]{article}
\input{psfig.sty}
\def\be{\begin{equation}}
\def\ee{\end{equation}}
\def\bea{\begin{eqnarray}}
\def\eea{\end{eqnarray}}

\begin{document}

\title{MULTIPOLE CORRECTIONS TO PERIHELION AND NODE LINE PRECESSION}
\author{ L. FERN\'ANDEZ-JAMBRINA }
\address{ETSI Navales, Universidad Polit\'ecnica de Madrid, \\ 
Arco de la Victoria s/n, \\
E-28040-Madrid\\E-mail: lfernandez@etsin.upm.es}

\maketitle\abstracts{In this talk relativistic corrections due to 
Geroch-Hansen multipoles for perihelion precession and node line precession 
of orbits in a stationary axially symmetric vacuum spacetime endowed 
with a plane of symmetry will be shown. 
Patterns of regularity will be discussed.}

\section{Introduction}

According to Kepler's law, a probe particle orbiting around a central 
spherical mass describes a trajectory with the shape of an ellipse. 
When particles are not pointlike or they are not even spherical, 
distortions appear and orbits are no longer closed, according to 
Bertrand's theorem. In such cases we talk of precession.

There are many phenomena comprised in this word, the precession of 
the rotation of the axis or the Earth, the precession of 
equinoxes\dots We shall deal with two of them.

First of all, we shall consider perihelion precession. In classical 
mechanics, it arises from deviations from sphericity, whereas in 
general relativity it comes up even for spherical bodies. 
Next we shall take into account the precession of the line of nodes of 
a nearly equatorial orbit.

\section{Mathematical framework}

For our purposes, we shall derive a stationary axially symmetric 
metric with a plane of symmetry. 
It is well known that Einstein equations in this case reduce to a 
second order partial differential equation for a complex function, 
the Ernst potential, $\varepsilon$,\cite{Ernst}

\begin{equation}
\varepsilon_{\rho\rho}+\frac{1}{\rho}\varepsilon_\rho+\varepsilon_{zz}=
\frac{2}{\varepsilon+\bar\varepsilon}({\varepsilon_\rho}^2+{\varepsilon_z}^2),
\label{ernst}\end{equation}\begin{equation} \varepsilon=f+i\chi,
\end{equation} where $f$ is a metric function and $\chi$ is the twist 
potential, that is non-zero for non-static metrics.

The metric components can be calculated from the Ernst potential as quadratures,
\begin{equation}
A_\rho=\frac{4\,\rho}{(\varepsilon+\bar\varepsilon)^2}\,\chi_z,\end{equation}
\begin{equation}
A_z=-\frac{4\,\rho}{(\varepsilon+\bar\varepsilon)^2}\,\chi_\rho,\label{a2}
\end{equation}

\begin{equation}
\gamma_\rho=\frac{\rho}{(\varepsilon+\bar\varepsilon)^2}
\,(\varepsilon_\rho\bar
\varepsilon_\rho-\varepsilon_z\bar\varepsilon_z),\label{g1} \end{equation}
\begin{equation}
\gamma_z=\frac{\rho}{(\varepsilon+\bar\varepsilon)^2}
\,(\varepsilon_\rho\bar
\varepsilon_z+\varepsilon_z\bar\varepsilon_\rho),\label{g2} \end{equation}

\begin{equation}
ds^2=-f(dt-Ad\phi)^2+\frac{1}{f}\{e^{2\gamma}(d\rho^2+dz^2)+\rho^2\,d\phi^2\}.
\end{equation}

Although the Ernst equation is an integrable system, we are very far 
from analitically implementing physics at will (Schwarzschild, Kerr 
solutions). Therefore, we shall work with aproximate solutions. We 
shall solve the Ernst equation up to the seventh order in the 
pseudospherical radius, $r=\sqrt{\rho^2+z^2}$, and the obtain the metric functions from it.

The results will be given in terms of invariant quantities. Concerning 
the probe, the existence of isometries leads to the appearance of two 
conserved quantities in geodesic motion, the energy and angular 
momentum per unit of mass of the test particle,
\begin{equation} E=-\partial_t\cdot u=f\,(\dot t-A\dot\phi),\end{equation}\begin{equation}
    l=\partial_\phi\cdot u=f\,A\,(\dot t-A\dot\phi)+\frac{1}{f}\,r^2\,\dot\phi, 
\end{equation} 
where the overhead dot stands for the derivative with respect to proper
time and $u$ is the velocity of the probe.

And considering the gravitational source, we have the Geroch-Hansen
multipole moments, $P_{n}$ \cite{Geroch}${}^,$\cite{Hansen}. In classical mechanics we would obtain the 
multipole moments as coefficients in the expansion of the 
gravitational potential in inverse powers of the radius and Legendre 
polynomials,\begin{equation} V(r,\theta)=-G\int 
d\vec{r}'\frac{\rho(\vec{r}')}{\|\vec{r}-\vec{r}'\|}=-G\sum_{n=0}^\infty 
P_{n}\frac{p_{n}(\cos\theta)}{r^{n+1}},\end{equation}
which can be easily read from the expansion on the axis,

\begin{equation} V(\rho=0)=-G\sum_{n=0}^\infty 
\frac{P_{n}}{z^{n+1}}.\end{equation}

In general relativity expressions are more involved, but still they 
can be obtained from the expansion of the Ernst potential in Weyl 
coordinates on the axis, \cite{Fodor}
\begin{equation}
    \varepsilon(\rho=0)=\sum_{n=0}^\infty\frac{C_{n}}{z^{n+1}},\end{equation}
\begin{equation} P_n=C_n,\quad n\leq 3,\end{equation}\begin{equation}
    P_4=C_4+\frac{1}{7}\,\bar C_0\,(C_1^2-C_2\,C_0)
\end{equation}
\begin{equation} P_5=C_5+\frac{1}{3}\,\bar
C_0\,(C_2\,C_1-C_3\,C_0)+\frac{1}{21}\bar C_1\,(C_1^2-C_2\, C_0),\dots
\end{equation}
although expressions remain simple only up to octupole moment.  Since 
we have imposed an equatorial plane of symmetry, even multipole 
moments will be real (gravitational moments) and odd ones will be 
imaginary (rotational moments). The reason for expanding up to seventh 
order is precisely the aim of reaching the first non-linear terms.

The expressions for the metric components are rather lengthy. Details 
will be published elsewhere.\cite{multi}

\section{Perihelion precession}

For perihelion precession the geodesic equations for equatorial orbits 
will be written in terms of the constants of motion. The Binet 
equation for the orbits,
\begin{equation}
{u_\phi}^2=e^{-2\gamma}\left\{\frac{E^2-f}{f^2\,(l-E\,A)^2}-u^2\right\}=F(u)-u^2,
\end{equation} is  equivalent to a quasilinear equation,
\begin{equation} u_{\phi\phi}=\frac{1}{2}\,F'(u)-u,
\end{equation}
which can be solved perturbatively, but a small parameter is needed. A good 
candidate is $\varepsilon=P_{0}/l$, since from Kepler's laws, 
$l\sim\sqrt{P_{0}r}$ in the far field region. To avoid secular terms a 
new coordinate, 
\begin{equation}
\psi=\omega\phi,\,\end{equation}
\begin{equation}\omega=\sqrt{1+\sum\omega_i\epsilon^i}.
\end{equation} 
is introduced and redefined at every step of perturbation. We are 
lead to a hierarchy of of harmonic equations for the terms of the 
series, 
\begin{equation}
u=\epsilon^2\,\sum_{n=0}^{11}u_n\,\epsilon^n+O(\epsilon^{14}), \end{equation}
the first of which is Keplerian.

But we are more interested in the frequency $\omega$, which will 
furnish the perihelion precession, 
\begin{eqnarray} {\Delta\phi}&=&2\,\pi\,(\omega^{-1}-1)=
\Delta_{class}+\Delta_{P_{0}}+\Delta_{P_{1}}+\Delta_{P_{2}}+
\Delta_{P_{3}}+
\nonumber\\&+&\Delta_{P_{4}}+\Delta_{P_{5}}+\Delta_{P_{1}-P_{2}}+
\Delta_{P_{1}-P_{3}}+\Delta_{P_{1}-P_{4}}+\Delta_{P_{2}-P_{3}}, \end{eqnarray} 
the terms of which we have classified here into classical and relativistic 
and these according to their multipole content.

Results are a bit cumbersome to produce here, but they can be 
summarized in the following way: 

The Schwarschild term, $\Delta_{P_{0}}$, as it is well 
known, is positive and the pure $P_{2}$ term is negative for positive 
quadrupole moment. This suggest a pattern of alternation of signs that 
has been checked up to the considered order of perturbation. For 
positive multipole moments, their linear contribution in 
$\Delta_{ P_{4n}}$ is positive and negative in $\Delta 
_{P_{4n+2}}$. This means, for instance, that for an 
approximately ellipsoidal source, all gravitational moment 
contributions are positive, since the sign of the multipole moments 
also alternates in the same way. There is no difference of sign 
between classical and relativistic terms.

On the other hand, rotational terms for $P_{2n+1}=iJ_{2n+1}$ are sensitive to the orientation 
of the orbit of the probe particle. For a counterrotating orbit the 
linear dipole term is positive, whereas the linear octupole term is 
negative. Again, there is a pattern of alternation of signs, which is 
similar to the one found for gravitational moments, the contribution 
of the pure multipole is positive for $\Delta_{P_{4n+1}}$ and negative 
for $\Delta_{P_{4n-1}}$ for a counterrotating orbit.

Finally the coupling terms bear the same sign as the product of the 
respective linear multipole terms, that is, $\textrm{sign\,}(\Delta_{P_{i}-P_{j}})= 
\textrm{sign\,}(\Delta_{P_{i}})\cdot\textrm{sign\,}(\Delta_{P_{j}})$. 
This rule is also true for self-couplings. Therefore all quadratic 
terms are positive.

\section{Line of nodes precession}

In this section we shall derive the precession of the nodes of a 
slightly tilted geodesic with respect to a nearby geodesic circle on 
the equatorial plane.

Perturbing the geodesic equations we arrive at and equation, 
\begin{equation}
\ddot\delta^\theta-\frac{1}{2}\,g^{\theta\theta}\,g_{ij,
\theta\theta}\,\dot x^i\,\dot x^j\,\delta^\theta=0, 
\end{equation}
\begin{equation}
\delta_{\phi\phi}+\Omega^2\,\delta=0,\end{equation}
\begin{equation}\Omega^2=-\left.\frac{1}{2\,
\dot\phi^2 }\, g^{\theta\theta}\,g_{\rho\sigma,\theta\theta}\right|_{r=R,\theta=\pi/2}
x^\rho\,\dot x^\sigma, \end{equation} 
for $\delta=\delta^\theta$, the zeros of which will determine the nodes of the orbit.

This equation can be handled perturbatively to yield the amount of 
node precession,
\begin{eqnarray} \Delta\phi&=&\Delta_{class}
+\Delta_{P_{1}}+\Delta_{P_{2}}
+\Delta_{P_{3}}+\Delta_{P_{4}}+\Delta_{P_{5}}+\nonumber\\&+&
\Delta_{ P_{1}-P_{2}}+\Delta_{P_{1}-P_{3}}+\Delta_{P_{1}-P_{4}}+
\Delta_{P_{2}-P_{3}},
\end{eqnarray}
 where there will be no contribution from the 
pure mass term, since for spherical distributions of mass there is no 
privileged plane of symmetry.

Concerning gravitational moments, the linear contribution of a 
positive quadrupole moment to $\Delta_{P_{2}}$ is positive. For 
higher multipoles the sign of the contributions alternates in the 
opposite way as for perihelion precession, that is, their linear contribution in 
$\Delta_{ P_{4n}}$ is negative and positive in $\Delta_{ P_{4n+2}}$.

A similar qualitative behaviour is found for rotational moments. The 
linear contribution of the dipole moment to $\Delta_{P_{1}}$ is 
negative for a counterrotating orbit. The pattern of signs is again 
the opposite to the one found for perihelion precession. The linear contribution 
of the  multipole is negative for $\Delta_{P_{4n+1}}$ and positive 
for $\Delta_{P_{4n-1}}$ for a counterrotating orbit.

So far, the main qualitative difference between both cases would be 
an overall sign. However new features appear when couplings are 
considered.

Classical terms follow the same pattern as the one derived for perihelion precession, $\textrm{sign\,}(\Delta_{P_{i}-P_{j}})= 
\textrm{sign\,}(\Delta_{P_{i}})\cdot\textrm{sign\,}(\Delta_{P_{j}})$, 
whereas the relativistic ones follow the opposite one, $\textrm{sign\,}(\Delta_{P_{i}-P_{j}})= 
-\textrm{sign\,}(\Delta_{P_{i}})\cdot\textrm{sign\,}(\Delta_{P_{j}})$. 

This behaviour has some curious consequences. Whereas  classical quadratic 
terms increase node precession, relativistic terms diminish it.

\section{Conclusions}

I would be interesting to explore whether the patterns observed for 
perihelion and node precession are general. However, there is no 
general formula for arbitrary orders of the multipole moments in terms 
of the Ernst potential expansion on the axis and this may pose severe 
difficulties. 

Other generalizations might include electromagnetic multipole moments 
and charged test particles or other precession phenomena, such as the 
dragging of gyroscopes.

\section*{Acknowledgments}
 The present work has been supported by Direcci\'on General de
Ense\~nanza Superior Project PB98-0772. L.F.J. wishes to thank
 F.J. Chinea, L.M. Gonz\'alez-Romero, F. Navarro-L\'erida and  M.J. Pareja 
 for valuable discussions.

\end{document}